\documentclass[10pt,twocolumn,letterpaper]{article}

\usepackage{iccv}
\usepackage{times}
\usepackage{epsfig}
\usepackage{graphicx}
\usepackage{amsmath}
\usepackage{amssymb}
\usepackage{multirow}
\usepackage[accsupp]{axessibility}

\usepackage[pagebackref=true,breaklinks=true,letterpaper=true,colorlinks,bookmarks=false]{hyperref}

\iccvfinalcopy 

\def\iccvPaperID{36} 

\ificcvfinal\fi

\begin{document}

\title{Classification and Visualization of \\ Genotype~$\times$~Phenotype Interactions in Biomass Sorghum}

\author{Abby Stylianou\\
\small Saint Louis University\\
{\tt\small astylianou@slu.edu}

\and
Robert Pless\\
\small George Washington University\\
{\tt\small pless@gwu.edu}

\and
Nadia Shakoor, Todd Mockler\\
\small Donald Danforth Plant Science Center\\
{\tt\small nshakoor|tmockler@danforthcenter.org}
}

\maketitle
\ificcvfinal\thispagestyle{empty}\fi

\begin{abstract}
We introduce a simple approach to understanding the relationship between single nucleotide polymorphisms (SNPs), or groups of related SNPs, and the phenotypes they control. The pipeline involves training deep convolutional neural networks (CNNs) to differentiate between images of plants with reference and alternate versions of various SNPs, and then using visualization approaches to highlight what the classification networks key on. We demonstrate the capacity of deep CNNs at performing this classification task, and show the utility of these visualizations on RGB imagery of biomass sorghum captured by the TERRA-REF gantry. We focus on several different genetic markers with known phenotypic expression, and discuss the possibilities of using this approach to uncover \textit{unknown} genotype~$\times$~phenotype relationships.
\end{abstract}

\section{Introduction}
Sorghum is a cereal crop, used worldwide for a variety of purposes including for use as grain and as a source of biomass for bio-energy production, which is the context we primarily focus on in this paper. For biofuel production, the goal of both plant growers and breeders is to produce sorghum crops that grow as big as possible, as quickly as possible, with as few resources as possible. Plant breeders produce new lines of sorghum by crossing together candidate lines that have desirable traits, or known genes that correspond to desirable traits.

In this paper, we propose a simple pipeline for understanding and identifying interesting genetic markers that control visually observable traits. This pipeline could be leveraged by plant geneticists and breeders to understand the relationship between single nucleotide polymorpishms (SNPs, locations in the organism's DNA that vary between different members of the population), or groups of related SNPs, and the phenotypes that they impact.

\vspace{.5cm}The pipeline involves:

\begin{itemize}
    \item Identifying candidate SNPs, or groups of SNPs, of interest in the sorghum genome;
    \item Training deep convolutional neural networks (CNNs) on visual sensor data to differentiate between reference and alternate versions of the SNP; and
    \item Visualizing what visual features led to a reference or alternate classification by the CNN.
\end{itemize}

We demonstrate the feasibility and utility of this pipeline on a number of SNPs identified in the sorghum Bioenergy Association Panel~\cite{brenton2016genomic} (BAP), a set of 390 sorghum cultivars whose genomes have been fully sequenced and which show promise for bio-energy usage.

\begin{figure}
    \centering
    \includegraphics[width=\columnwidth]{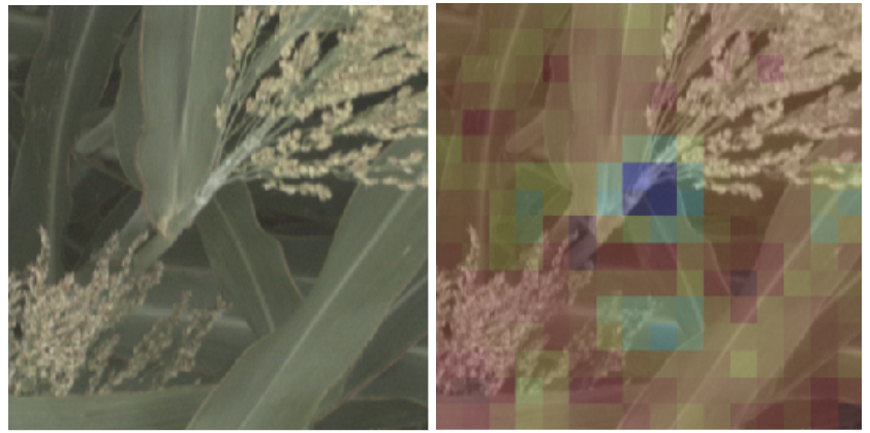}
    \caption{We train deep convolutional neural network classifiers to predict whether an image of a sorghum crop contains a reference or alternate version of particular genetic marker, and then visualize why the network makes that prediction. In this figure, we show the visualization for why the neural network predicted an image showed a plant with a reference version of a SNP that controls leaf wax composition -- the visualization highlights the especially waxy part of the stem.}
    \label{fig:front_page}
\end{figure}

\section{Background}
\subsection{Sorghum and Polymorphisms}
Sorghum is a diploid species, meaning that it has two copies of each of its 10 chromosomes. Each chromosome consists of DNA, the genetic instructions for the plant. The DNA itself is made up of individual nucleotides, sequences of which tell the plant precisely which proteins to make. Variations in these sequences, called single nucleotide polymorphisms, can result in changes to the proteins the plant is instructed to make, which in turn can have varying degrees of impact on the structure and performance of the plant. Understanding the impact that specific genes have on plants and how they interact with their environment is a fundamental problem and area of study in plant biology~\cite{bochner2003new, boyles2019genetic, cobb2013next, schweitzer2008plant}.

Single nucleotide polymorphisms (SNPs) are specific variations that exist between different members of a population at a single location on the chromosome, where one adenine, thymine, cytosine or guanine nucleotide in one plant may be have one or more different nucleotides in a different plant. This variation can exist on one or both copies of the chromosome. A cultivar that has the `original' version of the SNP on both copies of the chromosome is referred to as being homozygous reference; a cultivar that has variant on both copies of the chromosome is referred to as being homozygous alternate; and a cultivar that has one normal and one variant version of the SNP is called heterozygous. In this paper we consider only the homozygous cases, and how deep convolutional neural networks can be used to predict whether imagery of sorghum plants shows a plant with a reference or alternate version of a particular SNP or family of related SNPs.

\subsection{TERRA-REF}
\begin{figure}
    \centering
    \includegraphics[width=\columnwidth]{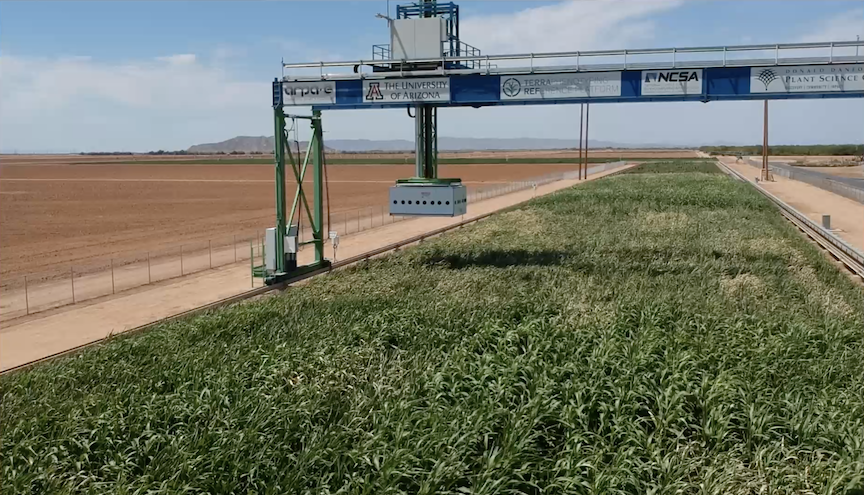}
    \caption[TERRA-REF Field Scanner]{The TERRA-REF Field and Gantry-based Field Scanner in Maricopa, Arizona, with sorghum being grown in the field.}
    \label{fig:field_scanner}
\end{figure}

We work with data collected by the Transportation Energy Resources from Renewable Agriculture Phenotyping Reference Platform, or TERRA-REF\cite{terra,LeBauer2020}, project which was funded by  the Advanced Research Project Agency--Energy (ARPA-E) in 2016. The TERRA-REF platform is a state-of-the-art gantry based system for monitoring the full growth cycle of over an acre of crops with a cutting-edge suite of imaging sensors, including stereo-RGB, thermal, short- and long-wave hyperspectral cameras, and laser 3D-scanner sensors. The goal of the TERRA-REF gantry was to perform in-field automated high throughput plant phenotyping, the process of making phenotypic measurements of the physical properties of plants at large scale and with high temporal resolution, for the purpose of better understanding the difference between crops and facilitating rapid plant breeding programs. The TERRA-REF field and gantry system are shown in Figure~\ref{fig:field_scanner}.

Since 2016, the TERRA-REF platform has collected petabytes of sensor data capturing the full growing cycle of sorghum plants from the sorghum Bioenergy Association Panel~\cite{brenton2016genomic}, a set of 390 sorghum cultivars whose genomes have been fully sequenced and which show promise for bio-energy usage. The full, original TERRA-REF dataset is a massive public domain agricultural dataset, with high spatial and temporal resolution across numerous sensors and seasons, and includes a variety of environmental data and extracted phenotypes in addition to the sensor data. More information about the dataset and access to it can be found in~\cite{LeBauer2020}.

\subsection{Deep Learning for Agriculture}
To our knowledge, this is the first work that trains classifiers on visual sensor data to predict whether an image shows organisms with a reference or alternate version of a genetic marker in order to better understand the genotype~$\times$~phenotype relationship. There is related work in genomic selection that attempts to predict end-of-season traits like leaf or grain length and crop yield~\cite{wheat_prediction, grain_yield_detection} from genetic information. In~\cite{liu2019phenotype}, the most related work to ours, the authors train CNNs to predict quantitative traits from SNPs, and use a visualization approach called saliency maps to highlight the \textit{SNPs} that most contributed to predicting a particular trait (as opposed to predicting whether a SNP is reference or alternate, and what visual components led to that classification). There is additionally work that attempts to use deep learning to predict the relative functional importance of specific genetic markers and mutations in plants~\cite{WANG202034}, without focusing on visualizing their specific impact on the expressed phenotypes.

There is generally significantly more work in applying deep learning for a wide variety of plant phenotyping and agriculture tasks that do not incorporate the underlying genetics -- for example, deep CNNs have successfully been used for fruit detection~\cite{wan2020faster,sa2016deepfruits,koirala2019deep,bargoti2017deep,lim2018durian}, cultivar and species identification~\cite{osako2020cultivar,heidary2021efficient,lim2018durian,ashqar2019plant,barre2017leafnet,van2018inaturalist,Ren2021MultiresolutionOP}, plant disease classification~\cite{wang2017automatic, ferentinos2018deep, mohanty2016using,too2019comparative, wang2017automatic}, leaf counting~\cite{ubbens2018use,aich2017leaf,giuffrida2018pheno,dobrescu2017leveraging}, yield prediction~\cite{maimaitijiang2020soybean, nevavuori2019crop,wang2018deep,chen2019strawberry}, and stress detection~\cite{chandel2021identifying,butte2021potato,anami2020deep}, among other phenotyping tasks.

\begin{figure*}[ht!]
    \centering
    \includegraphics[width=\textwidth]{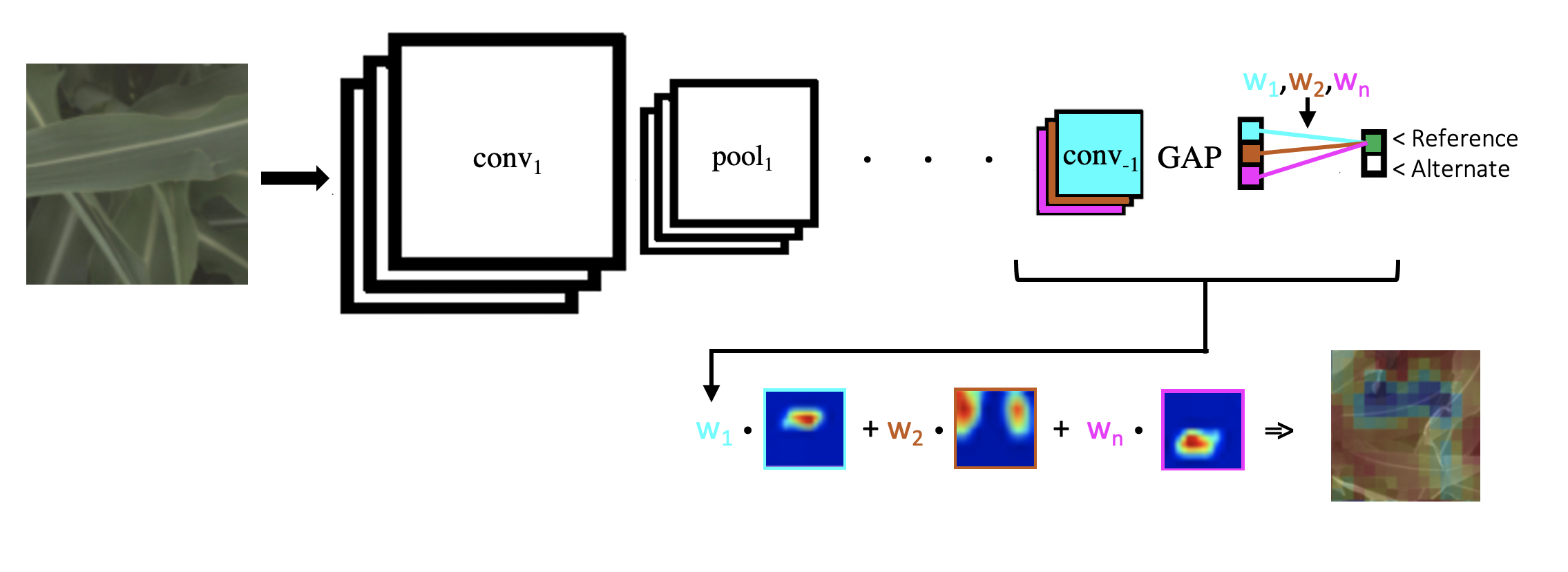}
    \caption{We use a standard ResNet-50 architecture, which like many deep convolutional neural networks consists of alternating convolutional and pooling layers (with interspersed activation functions). The network ends with a final convolutional layer, a global average pooling (GAP) operation, and then a fully connected layer, the output of which is used to make our prediction of whether an image shows a plant with a reference or alternate version of a particular genetic marker. We use the class activation mapping approach described in~\cite{zhou2015cnnlocalization}, in which the filters in the last convolutional layer are multiplied by the corresponding weights between the respective layer and the predicted output node. These weighted filters are then added up to produce a heatmap that has its highest values in important regions.}    \label{fig:architecture_and_cam}
\end{figure*}
\section{Methods}
Our approach to gaining understanding about the genotype~$\times$~phenotype relationship is to train deep convolutional neural networks to predict whether an image shows a sorghum cultivar with the reference or alternate version of a specific SNP or families of SNPs, and to then visualize what visual features the network focuses on when making that determination. If the classifier can perform well above chance at this classification task, then it is learning something that is significantly correlated with the genetics being considered, and the visualizations can help us glean insights into precisely what those correlations are.

In this paper, we focus on five specific families of genetic markers, as defined in Table~\ref{tab:snp_details}. Each famility of genetic markers is defined by one or more related SNPs, which have been identified in prior work as having a particular phenotype that is impacted depending on whether the cultivar being grown has the reference or alternate version of the marker. (When grouping multiple SNPs together, we consider a cultivar to be reference if it is reference for all of the SNPs, or alternate if it has any alternate SNPs -- this is because even one polymorphism can significantly impact the phenotype being controlled.)

We train a ResNet-50 deep convolutional neural network architecture~\cite{resnet} with a single fully connected layer on the reference vs. alternate classification task. A general overview of this type of network architecture is shown at the top of Figure~\ref{fig:architecture_and_cam}.

For all families of genetic markers, the network is trained on 256 $\times$ 256 pixel RGB images from the TERRA-REF Stereo Top RGB cameras, and optimized using adam~\cite{adam} with a learning rate of 0.0001 for 20 epochs. For data augmentation, we subtract by dataset channel means and divide by dataset channel standard deviations, and during training we perform random horizontal flips. The 256 $\times$ 256 pixel images extracted by resizing the image on its largest side to 256 and extracting a random crop at training time, and a center crop at testing time. We use imbalanced batch sampling during training to fill 100 image batches with a roughly equal number of reference and alternate images per batch, even if there is an imbalance in the number of reference and alternate images in the training set.

We then use the Class Activation Mapping approach described in~\cite{zhou2015cnnlocalization}, which highlights the image regions that most contributed to a classification of the neural network. This approach is visualized in the bottom of Figure~\ref{fig:architecture_and_cam}, where the filters in the last convolutional layer are multiplied by the corresponding weights between the respective layer and the predicted output node. These weighted filters are then added up to produce a heatmap that has its highest values in important regions (e.g., the blue regions in Figure~\ref{fig:front_page}). We can then look at the heatmaps for the most confident predictions from a neural network trained on a particular genetic marker family to understand the visual traits that are highly correlated with being either reference or alternate, as in Figures~\ref{fig:mostactivated1}~and~\ref{fig:mostactivated2}.

\begin{table*}[]
    \centering
    \resizebox{\textwidth}{!}{\begin{tabular}{|c|c|c|c|c|}
        \hline
        \multirow{2}{*}{\textbf{Genetic Marker Family}} & \multicolumn{4}{|c|}{\textbf{SNP Details}} \\
         & Chromosome & Gene & Position & Known Controlled Phenotype \\ \hline
        \multirow{4}{*}{Leaf Wax} & 1 & 001G269200 & 51588525 & \multirow{4}{*}{Wax composition~\cite{leaf_wax}}\\
                 & 1 & 001G269200 & 51588838 &\\
                 & 1 & 001G269200 & 51589143 &\\
                 & 1 & 001G269200 & 51589435 & \\ \hline
        \multirow{2}{*}{dw} & 6 & 006G067700 & 42805319 &\multirow{2}{*}{Plant height and structure, stem length and internode length~\cite{dw1,dw2}}\\
            & 6 & 006G067700 & 42804037 & \\ \hline
        \multirow{6}{*}{Dry Stalk (d) locus} & 6 & 006G147400 & 50898459 & \multirow{6}{*}{Plant height and structure, and sugar composition~\cite{dlocus}} \\
         & 6 & 006G147400 & 50898536 &\\
         & 6 & 006G147400 & 50898315 &\\
         & 6 & 006G147400 & 50898231 &\\
         & 6 & 006G147400 & 50898523 &\\
         & 6 & 006G147400 & 50898525 &\\ \hline
        \multirow{2}{*}{ma} & 6 & 006G057866 & 40312463 & \multirow{2}{*}{Flowering time and maturity~\cite{ma1, ma6}}\\
         & 6 & 006G004400 & 2697734 & \\  \hline
        tan & 9 & 009G229800 & 57040680 & Pigmentation and tannin production~\cite{tan1} \\ \hline
    \end{tabular}}
    \vspace{0.1cm}
    \caption{\textbf{Details about the genetic marker families of interest.} Single nucleotide polymorphisms are grouped by the phenotypes they control, and classification is performed by genetic marker family. Cultivars are defined as reference if they have the reference version of all SNPs on both copies of the chromosomes, and as alternate if they have the alternate version of all SNPs on both copies of the chromosomes (we do not consider heterozygous cultivars).}
    \label{tab:snp_details}
\end{table*}

\begin{table*}
    \centering
        \begin{tabular}{|c|cc|cc||cc|cc|}
        \hline
        \multirow{2}{*}{\textbf{Genetic Marker Family}} & \multicolumn{2}{|c|}{\textbf{\# Train Cultivars}} & \multicolumn{2}{|c||}{\textbf{\# Test Cultivars}} & \multicolumn{2}{c|}{\textbf{\# Train Images}} & \multicolumn{2}{|c|}{\textbf{\# Test Images}} \\
        & Ref & Alt & Ref & Alt & Ref & Alt & Ref & Alt \\ \hline
        Leaf Wax & 67 & 114 & 34 & 34 & 16750 & 28500 & 8500 & 8500 \\\hline
        dw & 80 & 105 & 40 & 40 & 20000 & 26250 & 10000 & 10000 \\\hline
        Dry Stalk (d) locus & 43 & 127 & 21 & 21 & 10750 & 31750 & 5200 & 5200 \\\hline
        ma & 21 & 167 & 10 & 10 & 5250 & 41750 & 2500 & 2500\\\hline
        tan & 133 & 53 & 27 & 27 & 33250 & 13250 & 6750 & 6750  \\ \hline
        \end{tabular}
    \vspace{0.1cm}
    \caption{\textbf{Dataset Statistics.} Specifics on the numbers of cultivars and images used in the training and testing sets for each of the genetic marker families.}
    \label{tab:dataset_details}
\end{table*}

\section{Dataset Details}
\label{sec:dataset}

For each family of genetic markers, we select the cultivars within the BAP lines that are either homozygous reference or alternate (ignoring heterozygous cultivars). We determine whether there are more reference or alternate cultivars, and select the minimum to define the number of cultivars that are put into our training and testing sets -- the testing set includes half of the cultivars from whichever class has fewer cultivars, and an equal number of the better represented class. There is no overlap between the training and testing cultivars.

Within each testing class, we randomly select the same number of images (the number is limited by whichever cultivar has the fewest images). This guarantees that our testing set is balanced both by number of images per class and number of cultivars per class. All remaining cultivars are put into the training set, without limiting the number of images per cultivar -- this allows us to use a large number of training examples, even if there may be imbalance in the number of images per class (reference vs. alternate) or per cultivar. This imbalance is dealt with at training time by an imbalanced sampler per batch, which selects roughly equal numbers of images from the population of reference and alternate examples. 

Table~\ref{tab:dataset_details} shows the exact number of cultivars and images used in the training and testing sets for each genetic marker family. We only consider images from June of 2017, mid-way through the growing season when plants are not too small, exhibiting many of the phenotypes of interest, and not yet lodging (falling over) on top of each other.

\section{Results}
In Table~\ref{tab:classification_acc}, we show the classification accuracy by image and by cultivar for each of the five genetic marker families of interest. The per image accuracy is simply the average accuracy of predicting whether every image in the test set is correctly labeled as homozygous reference or homozygous alternate. When computing the per cultivar accuracy, we take the mode of all images within a cultivar and use that to make the prediction. We then report the average accuracy over all cultivars. Due to the balancing described in Section~\ref{sec:dataset}, random chance on either of these tasks is guaranteed to be 0.5. For each of the genetic markers of interest, our models achieve well above chance accuracy, ranging from between 62.75\% accuracy for the dw marker and 68.59\% accuracy for the tan marker when considering each individual image. Taking the mode by cultivar provides an average improvement of nearly 20\% when compared with considering images individually.
\begin{table}
    \centering
    \begin{tabular}{|c|c|c|}
        \hline
        \multirow{2}{*}{\textbf{Genetic Marker Family}} & \multicolumn{2}{|c|}{\textbf{Classification Accuracy}} \\
         & Per Image & Per Cultivar \\ \hline
        Leaf Wax & .6325 & 0.7647 \\\hline
        dw & .6275 & 0.8375 \\\hline
        Dry Stalk (d) locus & .6743 & 0.8333 \\\hline
        ma & .6565 & 0.8500 \\\hline
        tan & .6859 & 0.8519 \\ \hline
    \end{tabular}\vspace{0.1cm}
    \caption{\textbf{Classification Accuracy by Image and by Cultivar.} Accuracy by image is computed on each image in the test set separately. Accuracy by cultivar is computed by taking the mode of the image predictions from each cultivar. The test set for each genetic marker family is balanced such that the classification accuracy by both image and by cultivar are 0.5.}
    \label{tab:classification_acc}
\end{table}

\begin{figure}
    \centering
    \includegraphics[width=\columnwidth]{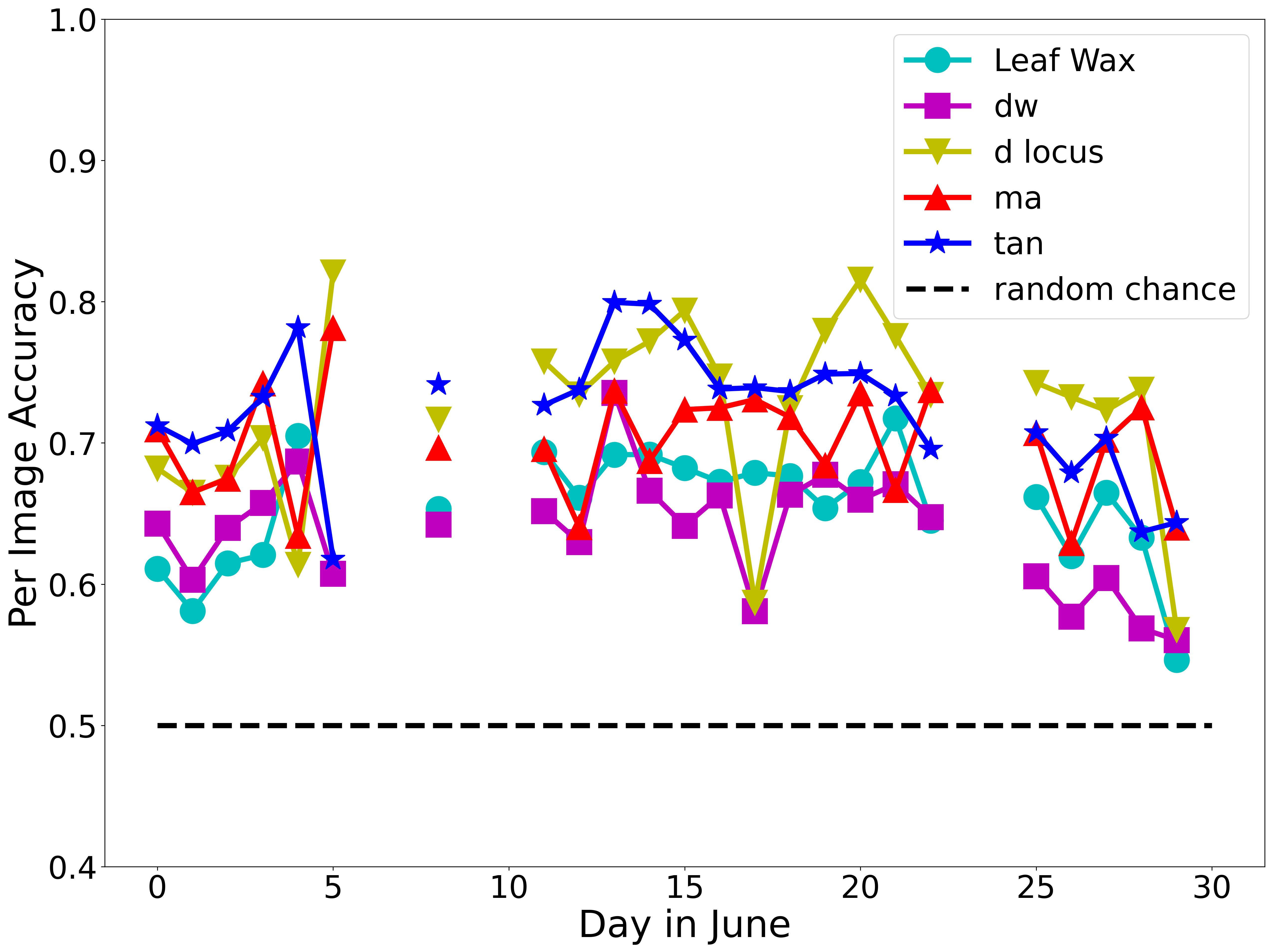}
    \caption{\textbf{Genetic Marker Family classification accuracy by date.} Here we show the accuracy per image for each genetic marker family on each day in June (mid-way through the growing season). In general performance is slightly better in the middle of the month, and worse at the end of the month when plants begin to lodge.}
    \label{fig:acc_by_date}
\end{figure}

We additionally show the accuracy on each day in June in Figure~\ref{fig:acc_by_date}. There is a slight trend across all of the markers showing slightly improved performance towards the middle of the month, with performance degrading significantly at the end of the month. This is likely explained by the phenotypes of interest being better expressed during this time and therefore being more recognizable, while the end of season degradation may be related to lodging that happens as the season progresses, where plants in adjacent plots start falling over into each other.

\begin{figure*}
    \centering
    \begin{tabular}{ccc}
        \multicolumn{3}{c}{(a) Leaf Wax} \\ 
        \raisebox{.4in}{\rotatebox{90}{Reference}} & \includegraphics[width=.65\columnwidth]{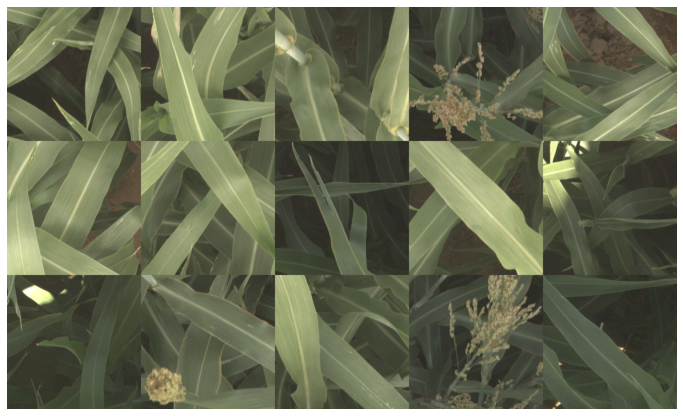} & \includegraphics[width=.65\columnwidth]{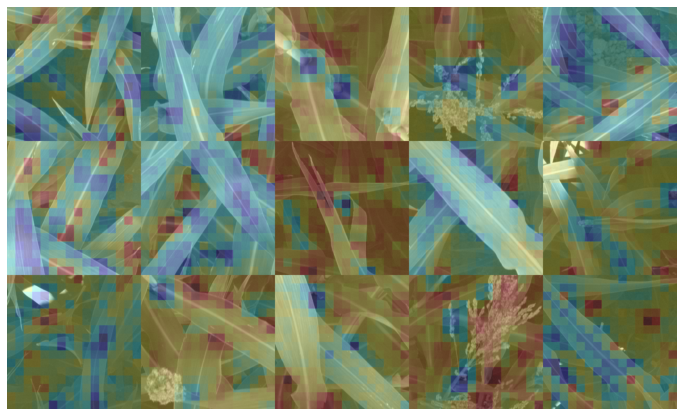} \\
        \raisebox{.4in}{\rotatebox{90}{Alternate}} & \includegraphics[width=.65\columnwidth]{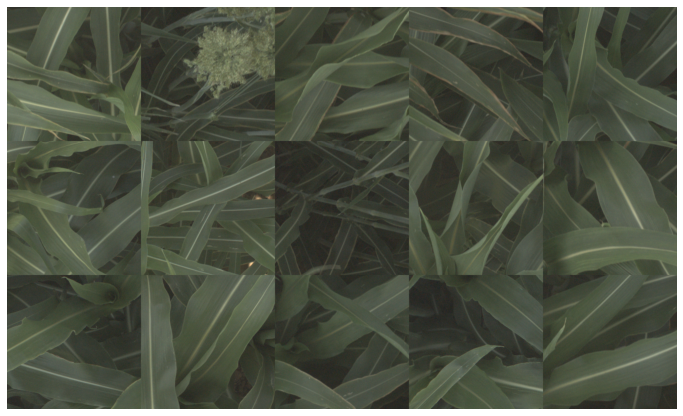} & \includegraphics[width=.65\columnwidth]{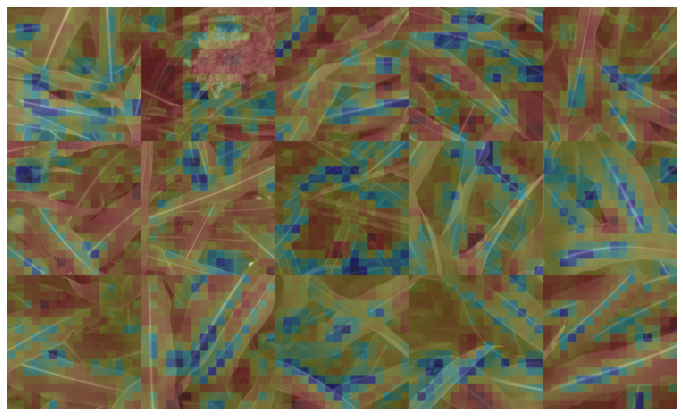} \\ 
        \hline
        
        \multicolumn{3}{c}{(b) dw} \\
        \raisebox{.4in}{\rotatebox{90}{Reference}} & \includegraphics[width=.65\columnwidth]{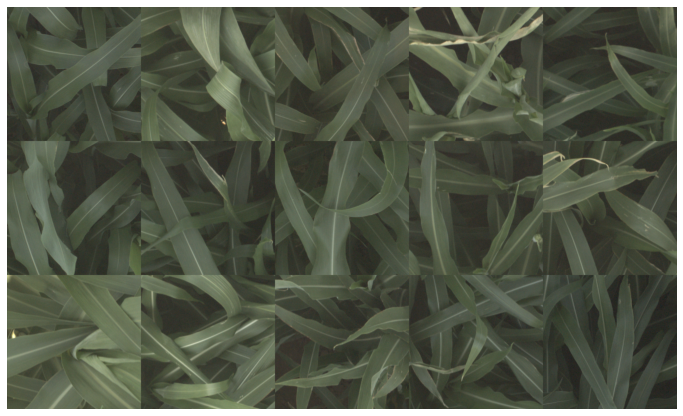} & \includegraphics[width=.65\columnwidth]{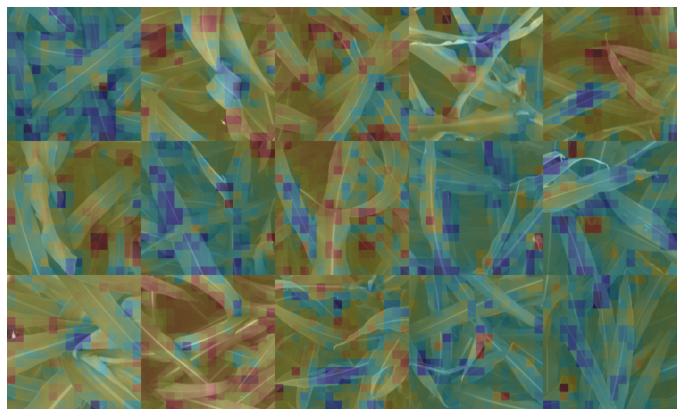} \\
        \raisebox{.4in}{\rotatebox{90}{Alternate}} & \includegraphics[width=.65\columnwidth]{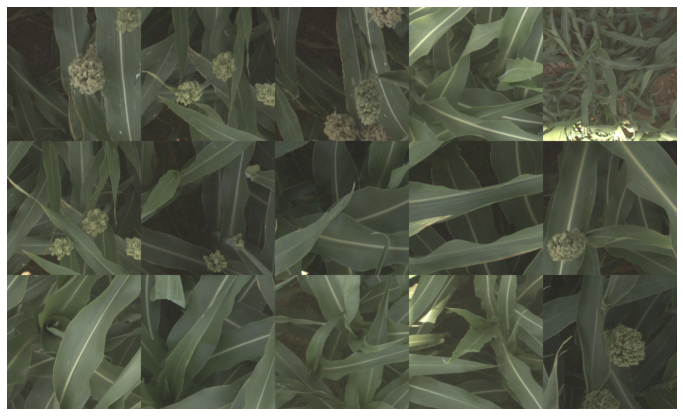} & \includegraphics[width=.65\columnwidth]{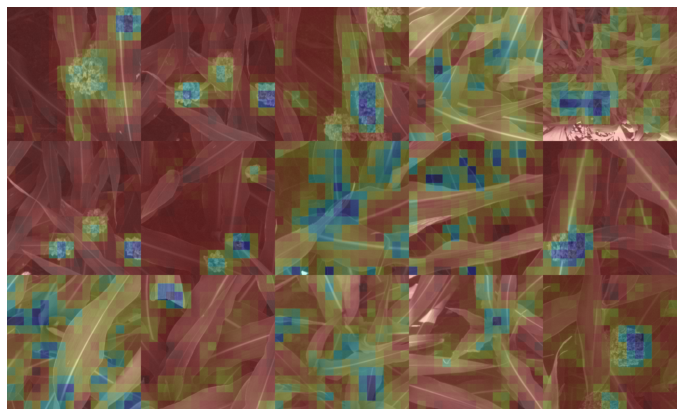} \\ 
        \\ \hline
        
        \multicolumn{3}{c}{(c) Dry Stalk (d) Locus} \\ 
        \raisebox{.4in}{\rotatebox{90}{Reference}} & \includegraphics[width=.65\columnwidth]{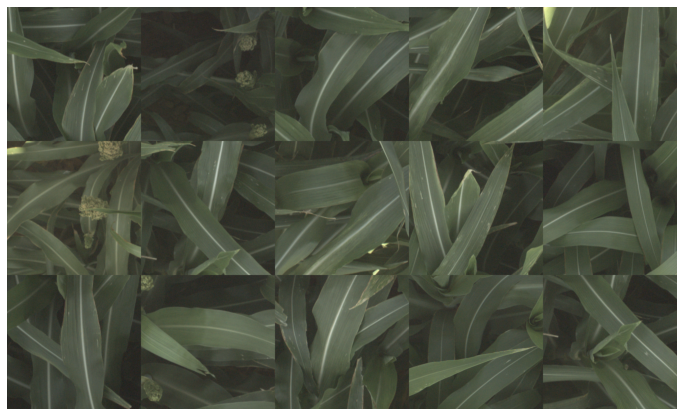} & \includegraphics[width=.65\columnwidth]{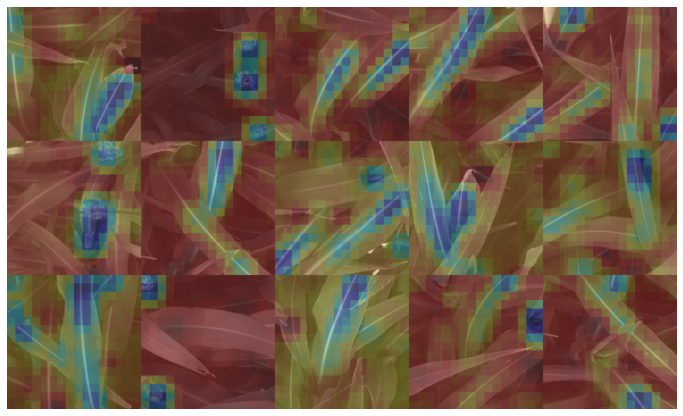} \\
        \raisebox{.4in}{\rotatebox{90}{Alternate}} & \includegraphics[width=.65\columnwidth]{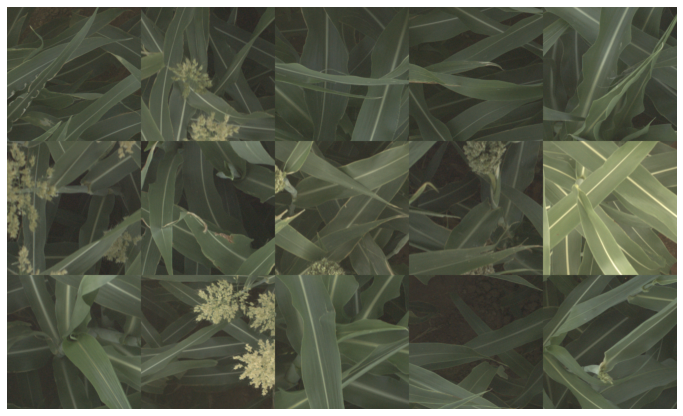} & \includegraphics[width=.65\columnwidth]{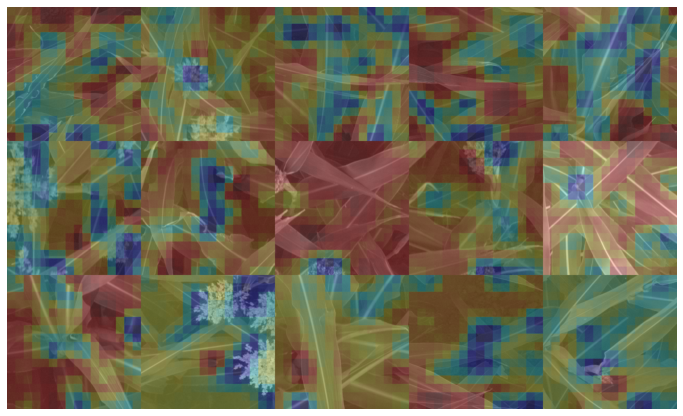} \\ 
    \end{tabular}
    \caption{Confident and correct examples for the classification of leaf wax, dw and d locus genetic markers. Blue regions are important to the prediction, while red regions are less important or even detract from the prediction.}
    \label{fig:mostactivated1}
\end{figure*}

\begin{figure*}
    \centering
    \begin{tabular}{ccc}
        \multicolumn{3}{c}{(d) ma} \\
        \raisebox{.4in}{\rotatebox{90}{Reference}} & \includegraphics[width=.65\columnwidth]{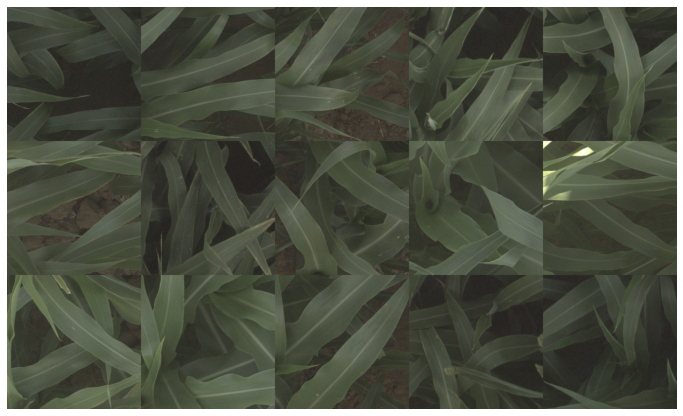} & \includegraphics[width=.65\columnwidth]{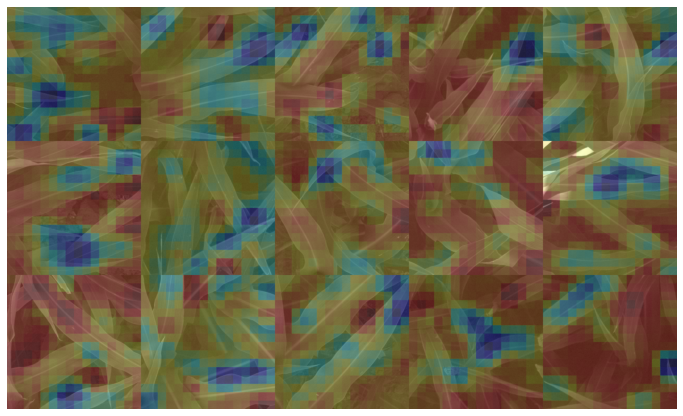} \\
        \raisebox{.4in}{\rotatebox{90}{Alternate}} & \includegraphics[width=.65\columnwidth]{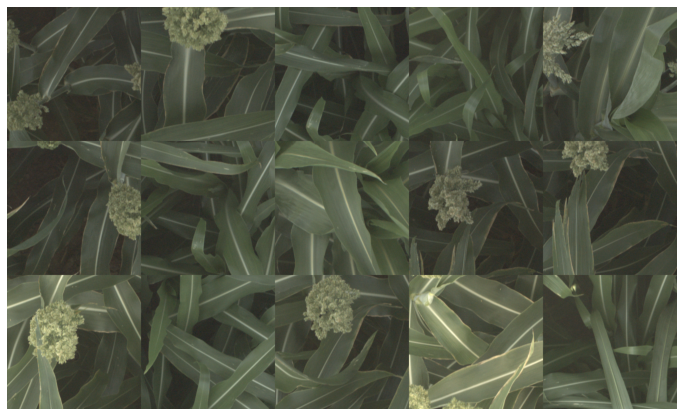} & \includegraphics[width=.65\columnwidth]{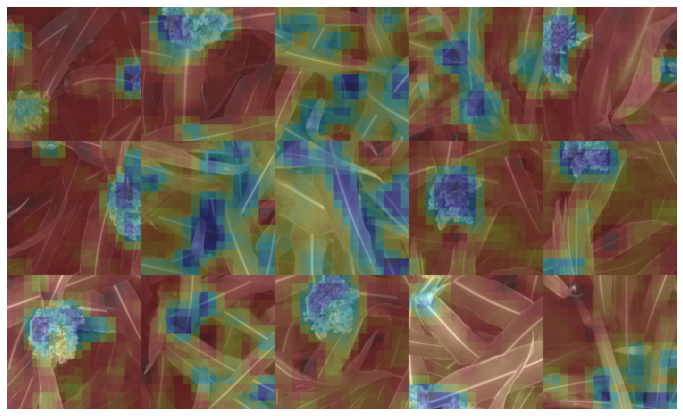} \\ 
        \\ \hline
        
        \multicolumn{3}{c}{(e) tan} \\
        \raisebox{.4in}{\rotatebox{90}{Reference}} & \includegraphics[width=.65\columnwidth]{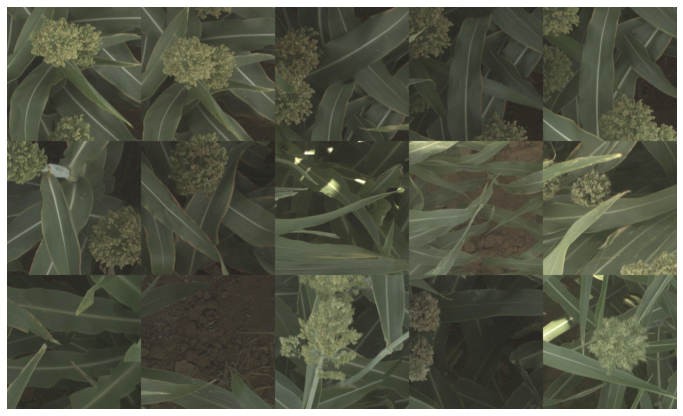} & \includegraphics[width=.65\columnwidth]{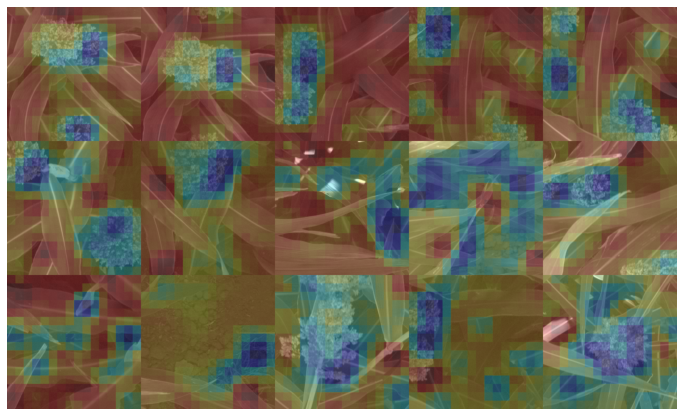} \\
        \raisebox{.4in}{\rotatebox{90}{Alternate}} & \includegraphics[width=.65\columnwidth]{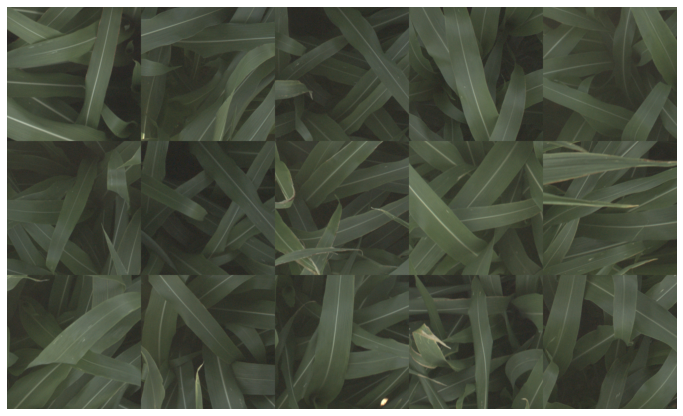} & \includegraphics[width=.65\columnwidth]{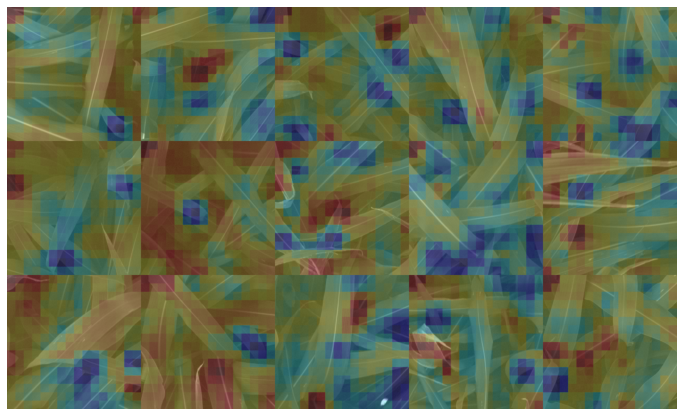} \\ 
        \hline
    \end{tabular}
    \caption{Confident and correct examples for the classification of ma and tan genetic markers. Blue regions are important to the prediction, while red regions are less important or even detract from the prediction.}
    \label{fig:mostactivated2}
\end{figure*}

In Figures~\ref{fig:mostactivated1}~and~\ref{fig:mostactivated2}, we show 15 of the most confident and correctly predicted reference and alternate images and their corresponding heatmaps for each of the genetic markers. These visualizations provide compelling insights into what the networks have learned to focus on, and therefore what visual plant features are highly correlated with a plant either being reference or alternate for a particular genetic marker. In the following paragraphs, we will discuss notable observations from these visualizations and how they correspond to the phenotypes these markers are known to control. In all visualizations, blue regions indicate visual features that are \textit{more important} for the classification, while red regions are less important.

\begin{figure}
\centering
\begin{tabular}{ccc}
     \includegraphics[width=.3\columnwidth]{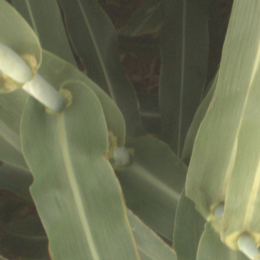} & \includegraphics[width=.3\columnwidth]{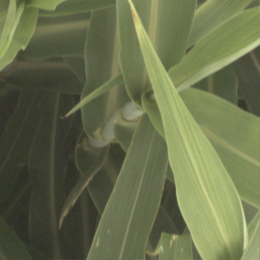} & \includegraphics[width=.3\columnwidth]{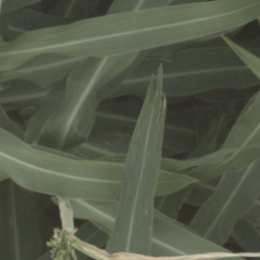}  \\
     \includegraphics[width=.3\columnwidth]{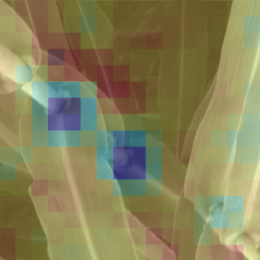} & \includegraphics[width=.3\columnwidth]{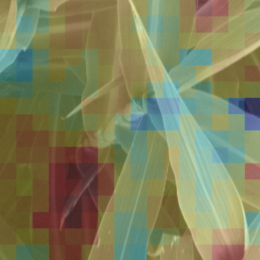} & \includegraphics[width=.3\columnwidth]{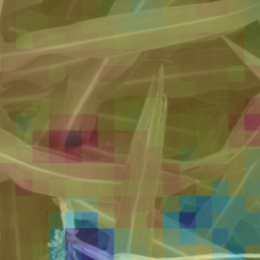} 
\end{tabular}
\caption{One of the features the neural network trained on the leaf wax SNPs learned to focus on when classifying an image as ``reference'' was the interface between the stem and either leaves or panicles. Field biologists who reviewed these visualizations were excited to see this feature highlighted, as this is often the feature they look at in the field as it's where wax build up is most obvious.}
\label{fig:leafwax}
\end{figure}

In Figure~\ref{fig:mostactivated1}(a), we show the most confident correct predictions for the leaf wax genetic marker family. Cultivars with the reference version of these SNPs are known to be more waxy, while the alternate versions are less waxy. In the reference heat maps, the important (blue) regions are often diffuse, covering much of the leaf, while the alternate visualizations are very focused on the spine of the leaf. Looking at the images, it is apparent that in the alternate images, this spine is more brightly differentiated from the rest of the leaf, while in the reference images the spine has less contrast. This corresponds to the wax build up on the leaf in the reference images, which cause the overall leaf to be whiter, resulting in lower contrast on the spine. When the reference visualizations are \textit{not} diffuse, they focus specifically on the interface between the sorghum plant spine and leaf -- this is shown in greater detail in Figure~\ref{fig:leafwax}. When reviewing these visualizations with a biologist on our team that does in-field ground truth phenotyping of traits including leaf wax, they said: ``That's exactly the place I look at when determining waxiness in the field -- it's where there's the wax is most obvious!'' Excitingly, this indicates that the network has learned, without explicit direction, to focus on the same plant parts as expert humans.

In Figures~\ref{fig:mostactivated1}(b and c) and~\ref{fig:mostactivated2}(d), the alternate visualizations appear to frequently focus on particular panicles at different growth stages (the panicles focused on for the dw and ma genetic markers are earlier in their life cycle when compared to the panicles in the d locus visualizations). This corresponds to the knowledge that polymorphisms in these genetic markers control features like plant growth rate (SNPs in the dw and dlocus families are considered `drawfing' markers, controling growth rate and ultimate plant height), flowering time and maturity. The d locus reference visualizations also appear to focus on particular leaf shapes -- the ends of broad leaves -- which similarly may relate to the fact that the markers are known to exhibit control over plant structure.

In Figure~\ref{fig:mostactivated2}(e), we show the confident images and heatmaps for the tan SNP, which controls tannin production and pigmentation. Notably, plants with the reference version of the SNP are known to have less tannin production, resulting in a less bitter taste in the plant seeds. This is interestingly manifested in the reference visualizations which highlight panicles where all of the seeds have been eaten -- likely because they taste better to birds!

\section{Conclusions \& Future Work}
In this paper, we demonstrated the feasibility of a pipeline to help understand the genotype~$\times$~phenotype relationship in sorghum by training deep convolutional neural networks on visual sensor data to predict whether different crops have reference or alternate versions of particular genetic markers. We show for several genetic markers that whose phenotypic expression is well understood that these networks can achieve well-above chance performance on this task, and that visualizations that highlight the most important parts of the images that led to the classification correspond with the known phenotypes.

This approach can be extended to not only help better understand well-established genotype~$\times$~phenotype relationships, but to explore new, less well understood relationships. The same approach could be deployed for SNPs and families of SNPs whose phenotypic expression is \textit{not} understood, to uncover new and interesting polymorphisms.

This paper presented a very simple, yet effective pipeline, focused on a relatively limited time period of high resolution data from the TERRA-REF gantry system (data from the entire month of June, mid-way through the growing season in 2017). We recognize that not all phenotypes, however, are observable during this time period. Especially when considering unknown genetic markers, it may be beneficial to consider longer time periods including both early and late growing periods when different phenotypes are expressed. This is a direction for future work: longer time periods may require more complex training protocols that more explicitly incorporate time -- for example, using recurrent approaches, or training a multi-headed network that simultaneously predicts the genetic class and the date. Additional work could focus on extending the approach to sensors other than RGB cameras, as some phenotypes may be more readily observed in different sensing modalities, such as hyperspectral or thermal imagery, or in the structural information from the 3D laser scanner.

\newpage
{\small
\bibliographystyle{ieee_fullname}
\bibliography{main}
}

\end{document}